\documentclass[aip,reprint,superscriptaddress,amsmath,amssymb,twocolumn,preprintnumbers]{revtex4-1}

\usepackage{graphics,graphicx}

\begin{document}
\preprint{Physics of Fluids}

\title{Unlikely existence of $k_x^{-1}$ spectral law in wall turbulence: an observation of the atmospheric surface layer
}

\author{Hideaki Mouri}
\affiliation{Meteorological Research Institute, Nagamine, Tsukuba 305-0052, Japan}
\affiliation{Graduate School of Science, Kobe University, Rokkodai, Kobe 657-8501, Japan}
\author{Takeshi Morinaga}
\affiliation{Meteorological Research Institute, Nagamine, Tsukuba 305-0052, Japan}
\author{Shigenori Haginoya}
\affiliation{Meteorological Research Institute, Nagamine, Tsukuba 305-0052, Japan}

\begin{abstract}
For wall turbulence, there has been predicted a range of streamwise wavenumbers $k_x$ such that the spectral density of streamwise velocity fluctuations is proportional to $k_x^{-1}$. The existence or nonexistence of this $k_x^{-1}$ law is examined here. We observe the atmospheric surface layer over several months, select suitable data, and use them to synthesize the energy spectrum that would represent wall turbulence at a very high Reynolds number. The result is not consistent with the $k_x^{-1}$ law. It is rather consistent with a recent correction to the prediction of a model of energy-containing eddies that are attached to the wall. The reason for these findings is discussed mathematically.

\end{abstract}

\maketitle

\section{Introduction} \label{S1}

Despite being fundamental to any turbulent flow, the overall shape of the energy spectrum is still controversial. This is true even for representative flows. Among them, we study the case of wall turbulence of an incompressible fluid, e.g., a boundary layer over a flat surface or a flow within a pipe.

To be specific, the $x$--$y$ plane is taken at the wall. The $x$ axis is aligned with the mean stream. While $U(z)$ denotes the mean velocity at a distance $z$ from the wall, $u(z)$ and $w(z)$ denote velocity fluctuations in the streamwise and wall-normal directions. The wall turbulence is assumed to be homogeneous in the $x$ and $y$ directions. Its total thickness $\delta$ remains some constant. We also assume that the turbulence is stationary.

Asymptotically in the limit of high Reynolds number, wall turbulence has a sublayer at $z /\delta \rightarrow 0$ such that its momentum flux $\rho \langle -uw \rangle$ takes a constant value of $\rho u_{\ast}^2$. Here $\rho$ is the mass density, $u_{\ast}$ is the friction velocity, and $\langle \cdot \rangle$ denotes an average. This constant-flux sublayer is just where $U$ varies logarithmically with $z$.\cite{my71,g92} At a finite but sufficiently high Reynolds number, it yet serves as a good approximation for a finite range of distances $z$.

The energy spectra in the constant-flux sublayer have been modelled by Perry and his collaborators.\cite{pa77,phc86} As for the spectral density of streamwise velocity fluctuations ${\mit\Phi}_u (k_x,z)$ at a streamwise wavenumber $k_x$, it is assumed that low-wavenumber modes are determined by $k_x$, $\delta$, and $u_{\ast}$ whereas high-wavenumber modes are determined by $k_x$, $z$, and $u_{\ast}$. These two are assumed to overlap from $k_x = a_1/\delta$ to $a_2/z$. Here $a_1$ and $a_2$ are constants of $\mathcal{O}(1)$. By ignoring the highest wavenumbers at which the fluid viscosity $\nu$ is also important,\cite{pa77}
\begin{subequations}
\label{eq1}
\begin{equation}
\label{eq1a}
{\everymath{\displaystyle}
{\mit\Phi}_u(k_x,z) = \left\{ \begin{array}{ll}
                                                     \frac{u_{\ast}^2}    {k_x} \, f_u(k_x \delta)  &\ \mbox{at} \                        k_x < \frac{a_1}{\delta} ,  \\
                                  \rule{0ex}{4.5ex}  \frac{c_u u_{\ast}^2}{k_x}                     &\ \mbox{at} \ \frac{a_1}{\delta} \le k_x < \frac{a_2}{z} ,       \\
                                  \rule{0ex}{4.5ex}  \frac{u_{\ast}^2}    {k_x} \, g_u (k_x z)      &\ \mbox{at} \ \frac{a_2}{z}      \le k_x  .
                              \end{array}
                      \right.
}
\end{equation}
While $f_u$ and $g_u$ are functions, $c_u$ is a constant. The law of ${\mit\Phi}_u \varpropto u_{\ast}^2/k_x$ in that overlapping range from $k_x = a_1/\delta$ to $a_2/z$ is known as the $k_x^{-1}$ law.\cite{t53}

This $k_x^{-1}$ law is not allowed for the spectral density of wall-normal velocity fluctuations ${\mit\Phi}_w (k_x,z)$.\cite{phc86} It has to be in the functional form of
\begin{equation}
\label{eq1b}
{\mit\Phi}_w(k_x,z) = \frac{u_{\ast}^2}{k_x} \, g_w (k_x z).
\end{equation}
\end{subequations}
Since $w$ is blocked by the wall, $w$ at this distance $z$ is due only to eddies of wall-normal sizes of $\mathcal{O}(z)$. There is no mode that reflects the total thickness $\delta$.

By integrating ${\mit\Phi}_u(k_x,z)$ and ${\mit\Phi}_w(k_x,z)$ in Eq.~(\ref{eq1}) over the whole range of wavenumbers $k_x$, the variances $\langle u^2(z) \rangle$ and $\langle w^2(z) \rangle$ are derived respectively as\cite{pa77,phc86}
\begin{subequations}
\label{eq2}
\begin{equation}
\label{eq2a}
\frac{\langle u^2(z) \rangle}{u_{\ast}^2} = b_u + c_u \ln \left( \frac{\delta}{z} \right),
\end{equation}
with $b_u = \int^{\infty}_{a_2}\!\!s^{-1}g_u(s)ds + c_u \ln (a_2/a_1) + \int^{a_1}_0\!\!s^{-1}f_u(s)ds$ and 
\begin{equation}
\label{eq2b}
\frac{\langle w^2(z) \rangle}{u_{\ast}^2} = b_w,
\end{equation}
\end{subequations}
with $b_w = \int^{\infty}_0\!\!s^{-1}g_w(s)ds$. For $\langle u^2 \rangle$ in Eq.~(\ref{eq2a}), the $k_x^{-1}$ law of Eq.~(\ref{eq1a}) has led to the logarithmic factor $\ln (\delta /z)$. With an increase in the ratio $\delta/z$, the overlapping range from $k_x = a_1/\delta$ to $a_2/z$ is increasingly broad and energy-containing. The variance $\langle u^2 \rangle$ is increasingly large.

The logarithmic law of Eq.~(\ref{eq2a}) has been established recently by means of laboratory experiments and field observations.\cite{mmhs13,hvbs12,hvbs13} Its constant $c_u$ appears to take a common value of about $1.3$.\cite{mmhs13,hvbs12,hvbs13,mm13,vhs15,ofsbta17,mmym17}

However, the $k_x^{-1}$ law itself has not been established so far. To derive Eq.~(\ref{eq2a}), another model is known, i.e., the attached-eddy hypothesis of Townsend. \cite{t76} Here, energy-containing motions of wall turbulence are attributed to a random superposition of eddies that are attached to the wall, are of various finite sizes, and have some common shape with the same characteristic velocity $u_{\ast}$. Although they had been expected to lead to the $k_x^{-1}$ law,\cite{phc86} such an expectation is not correct.\cite{m17} For the attached eddies, the mathematically correct form of ${\mit\Phi}_u$ is
\begin{equation}
\label{eq3}
{\mit\Phi}_u(k_x,z) = \frac{u_{\ast}^2}{k_x} \left[ g_{u:1}(k_x z) + \ln \left( \frac{\delta}{z} \right) g_{u:2}(k_x z) \right].
\end{equation}
At any of $k_x$, Eq.~(\ref{eq3}) depends on $\delta$. The largest eddies of a finite streamwise size of $\varpropto \delta$ affect all the wavenumbers $k_x$, since any spatial fluctuations and their Fourier transforms do not simultaneously have compact supports,\cite{s66,b85} i.e., finite ranges only where the function is nonzero (see Sec.~\ref{S6a}). With an increase in the ratio $\delta/z$, the factor $\ln (\delta/z)$ in Eq.~(\ref{eq3}) is increasingly important. The variance $\langle u^2 \rangle$ is increasingly large, with $b_u = \int^{\infty}_0\!\!s^{-1}g_{u:1}(s)ds$ and $c_u = \int^{\infty}_0\!\!s^{-1}g_{u:2}(s)ds$ as for Eq.~(\ref{eq2a}). On the other hand, the form of ${\mit\Phi}_w$ is the same as in Eq.~(\ref{eq1b}).\cite{m17}

\begin{figure}[tbp]
\resizebox{7.5cm}{!}{\includegraphics*[4.2cm,18.7cm][18.9cm,28.4cm]{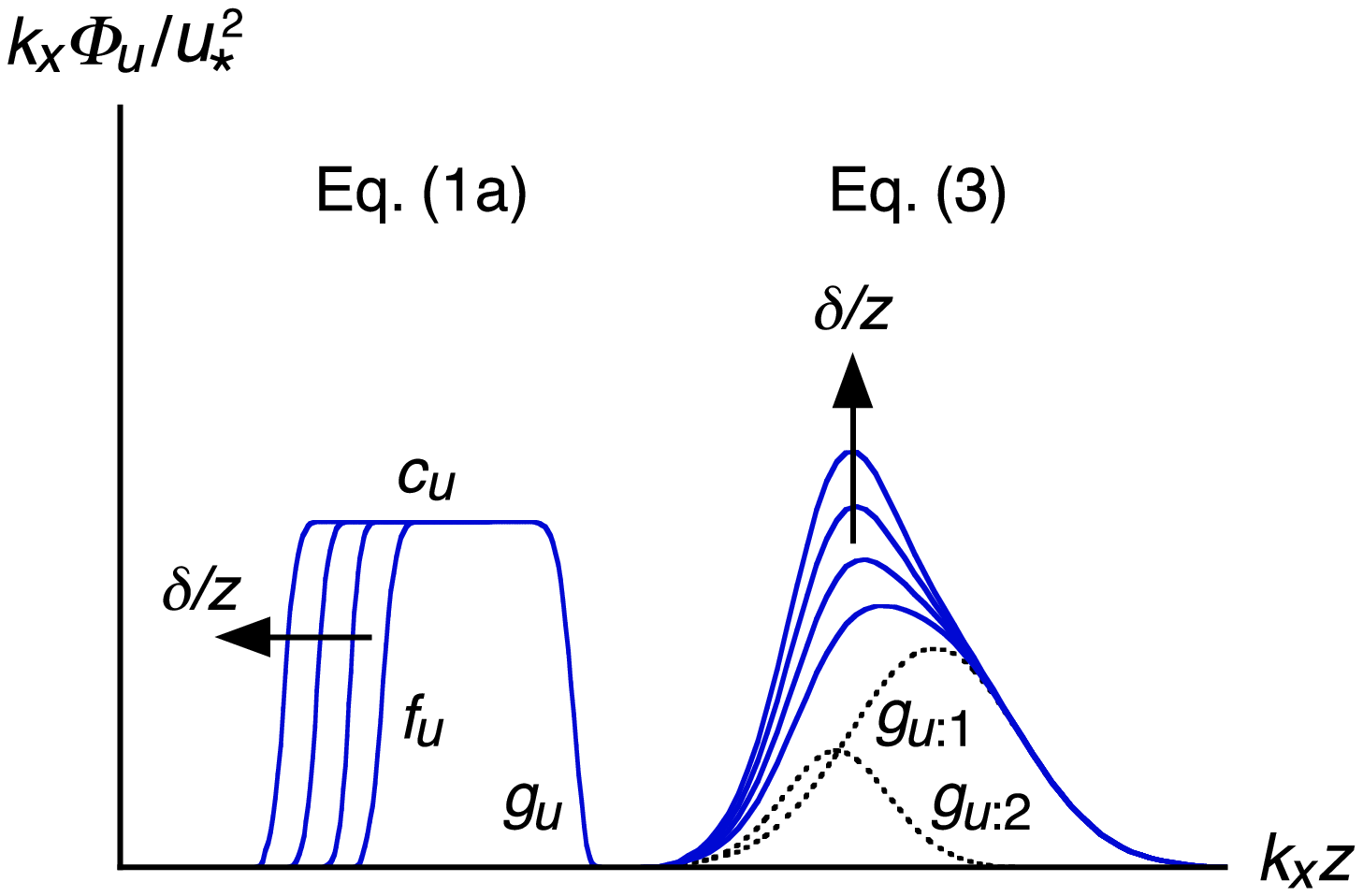}}
\caption{\label{f1} Schematic of the premultiplied spectrum $k_x {\mit\Phi}_u/u_{\ast}^2$ in Eqs.~(\ref{eq1a}) and (\ref{eq3}) as a function of $k_x z$. The arrows indicate increasing $\delta /z$. As for Eq.~(\ref{eq3}), the dotted lines denote the functions $g_{u:1}$ and $g_{u:2}$.}
\end{figure}

Thus, it is required to examine the very shape of the energy spectrum in a laboratory experiment or in a field observation. Any direct numerical simulation is not considered here, since its Reynolds number is still not high enough to reproduce the logarithmic law of Eq.~(\ref{eq2a}).\cite{lm15} Following the recent studies,\cite{ff99,hhs02,dckbfr04,zs07,rhvbs13,km06,vgs15,wz16} we focus on the premultiplied spectrum $k_x {\mit\Phi}_u$ so that the $k_x^{-1}$ law  of Eq.~(\ref{eq1a}) would show up as a plateau from $k_x =a_1/\delta$ to $a_2/z$ as sketched in Fig.~\ref{f1}.

Even in recent experiments of pipe flows and boundary layers at high Reynolds numbers, while $\langle u^2 \rangle$ is logarithmic over distances $z$ from the wall,\cite{hvbs12,hvbs13,vhs15} the existence of the $k_x^{-1}$ law remains controversial at those distances.\cite{zs07,rhvbs13,vgs15} Instead of an exact plateau of $k_x {\mit\Phi}_u/u_{\ast}^2 = c_u \simeq 1.3$, there is a peak at $k_x z \simeq$ several times $10^{-2}$. With an increase in the ratio $\delta /z$, the peak is increasingly high as expected rather in Eq.~(\ref{eq3}) that has been derived from the attached-eddy hypothesis (see Fig.~\ref{f1}).

The plateau of $k_x {\mit\Phi}_u$ is yet found in some observations of the atmospheric surface layer,\cite{ff99,hhs02,dckbfr04} i.e., the constant-flux sublayer of the atmosphere at $z \lesssim 100$\,m where the Reynolds number is higher and the ratio $\delta /z$ is larger than those in the existing experiments.\cite{g92} Another observation has captured the $k_x^{-1}$ law of ${\mit\Phi}_u$ over more than a decade and a half in wavenumber $k_x$.\cite{chosp13}

These features are not found in other observations.\cite{km06,wz16} For each of such field observations,\cite{ff99,hhs02,dckbfr04,km06,chosp13,wz16} the shape of ${\mit\Phi}_u$ is unlikely to have converged in a statistical sense.\cite{hhs02,km06} Their durations are not long, i.e., $30$ to $90$\,min, even under a mean wind velocity of only a few times $10^0$\,m\,s$^{-1}$. It is not avoidable in the atmospheric surface layer. Over a longer duration, the mean wind velocity and the mean wind direction would vary largely.

We are to obtain a representative shape of the premultiplied spectrum $k_x {\mit\Phi}_u$ in the atmospheric surface layer. The observation had been continued over several months (Sec.~\ref{S2}), from which we select data that are suited to our study (Sec. \ref{S3}). On the basis of the same assumption as for the $k_x^{-1}$ law of Eq.~(\ref{eq1a}), these data are used to synthesize the spectral density ${\mit\Phi}_u$ at each $k_x$ (Sec.~\ref{S4}). Having found that the resultant shape of $k_x {\mit\Phi}_u$ is not consistent with the $k_x^{-1}$ law (Sec.~\ref{S5}), we discuss the mathematical reason and so on in Secs.~\ref{S6} and \ref{S7}.

\begin{figure}[bp]
\resizebox{8.7cm}{!}{\includegraphics*[1.0cm,10.0cm][20.5cm,20cm]{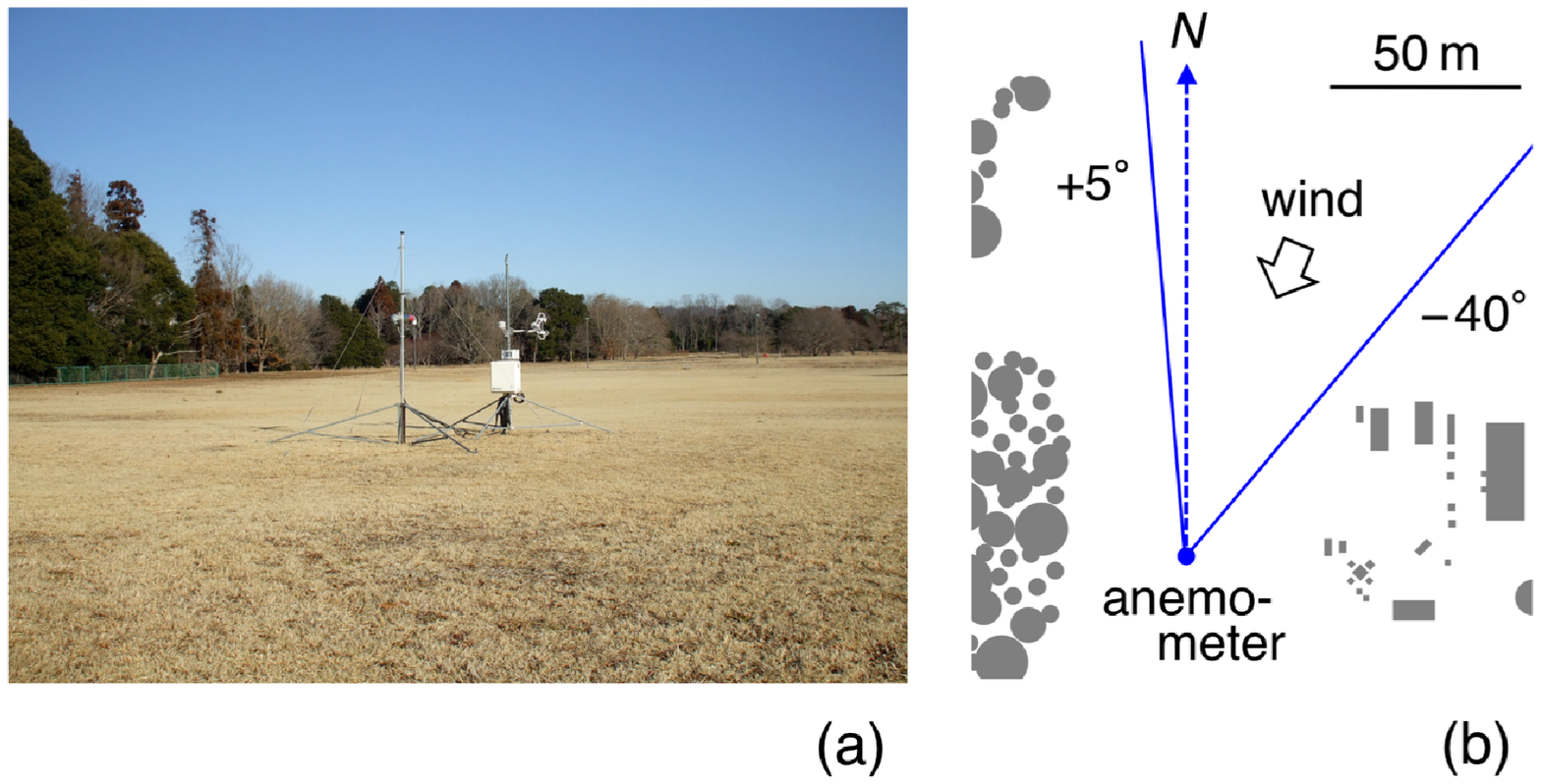}}
\caption{\label{f2} (a) Southeast view of the anemometer. On its left, there is a mast equipped with a reference thermometer. (b) Top-view schematic of the site. The gray areas denote obstacles such as woods, huts, and measuring devices.}
\end{figure} 

\section{Observation} \label{S2}

The observation had been continued from $1$ December $2016$ to $10$ April $2017$ and from $21$ November $2017$ to $31$ March $2018$ at a flat grass field within the Meteorological Research Institute ($36.055^{\circ}$N, $140.123^{\circ}$E). This field is $200$\,m in the north-south direction and $100$\,m in the west-east direction. As shown in Fig.~\ref{f2}(a), the grass was dry and had been cut to heights of $\lesssim 50$\,mm just before each of the two observing periods.

Because of woods lying along the western border of the observing field (Fig.~\ref{f2}), we are to study cases of north winds that were prevailing during those two periods. Up to a windward distance of 100\,m from the northern border, the surface condition remains almost the same. From 100 to 300\,m, there lie occasional trees with heights of $\lesssim 10$\,m. Beyond 300\,m, a wooded area continues.

These conditions are not optimal. Nevertheless, since the field is within our institute, we were able to maintain continuously the measuring devices. Such a maintenance is crucial to any long observation. In addition, the field is adjacent to an observing site of the Japan Meteorological Agency. Its routine observations are used in Secs.~\ref{S3} and \ref{S6d} to select and justify our data.

To measure the wind, we installed a triaxial ultrasonic anemometer (Campbell, CSAT3), pointed to the north, at the center of the observing field (Fig.~\ref{f2}). The height from the surface was $z = 1.75$\,m. It is low enough to have a large value of $\delta/z$ and is high enough to be not affected directly by the surface roughness (see Sec.~\ref{S3}). With a spatial resolution of $100$\,mm in the vertical direction and $58$\,mm in the horizontal direction, this anemometer measures all the three components of the wind velocity along with the average ${\mit\Theta}$ and fluctuations $\theta$ of the air temperature. The measurement errors are $\lesssim 2$\,\% under conditions such as of ours.

The data are made of many subsets. For each of them, the duration was $30$\,min. The sampling frequency $f_{\rm s}$ was $10$\,Hz. Out of the resultant 18,000 records per quantity, $2^{14} = 16,384$ records from the first are to be used in the following calculations.

\section{Data Selection} \label{S3}

Before selecting data that are suited to our study, the mean wind direction is determined in each of the data subsets as
\begin{subequations}
\label{eq4}
\begin{equation}
\label{eq4a}
\alpha = \tan^{-1} \left( \frac{\langle v_{{\rm we}} \rangle}{\langle v_{{\rm ns}} \rangle} \right) .
\end{equation}
Here $v_{{\rm we}}$ and $v_{{\rm ns}}$ are instantaneous velocities, measured by the anemometer, in the west-east and in the north-south directions. Then, 
\begin{equation}
\begin{pmatrix}
U+u \\
v
\end{pmatrix}
=
\begin{pmatrix}
 \cos \alpha & \sin \alpha \\
-\sin \alpha & \cos \alpha
\end{pmatrix}
\begin{pmatrix}
v_{{\rm ns}} \\
v_{{\rm we}}
\end{pmatrix}
.
\end{equation}
\end{subequations}
A similar conversion is applied to the vertical velocity $w$ such that its average is equal to $0$. The friction velocity $u_{\ast}$ is subsequently estimated as $\langle -uw \rangle^{1/2}$.

To avoid any effect of obstacles such as woods around our observing field (Fig.~\ref{f2}), we only use subsets with the mean wind direction $\alpha$ from $-40^{\circ}$ to $+5^{\circ}$, i.e., from the northeast to the north-northwest directions. The mean wind velocity $U$ is limited to a range from $1.5$ to $4.5$\,m\,s$^{-1}$ so that the final data would be as homogeneous as possible.

There are cases of precipitation, e.g., rain, which affects turbulence in the atmosphere. We exclude them by using routine data of the Aerological Observatory of the Japan Meteorological Agency located at $500$\,m from our observing field.

We focus on the near-neutral cases that do not suffer significantly from buoyancy due to a vertical variation of the temperature ${\mit\Theta}+\theta$. Since the Monin-Obukhov length $L_{\ast}$ diverges in an exactly neutral case,\cite{my71,g92} its value at our observing height $z = 1.75$\,m is used to impose a criterion that has to be satisfied by any of the data subsets studied hereafter,
\begin{subequations}
\label{eq5}
\begin{equation}
\label{eq5a}
\left\vert \frac{z}{L_{\ast}(z)} \right\vert < 0.1
\ \
\mbox{for}
\
L_{\ast}(z) = \frac{u_{\ast}^3 {\mit\Theta}(z)}{\kappa g  \langle -\theta w(z) \rangle} .
\end{equation}
While $\kappa = 0.4$ is the von K\'arm\'an constant, $g$ is the local gravitational acceleration. The threshold of $0.1$ is from the previous studies.\cite{hhs02,km06,kc98}

To exclude cases where the wind direction is not stationary and has varied too largely, we impose a criterion as\cite{kc98}
\begin{equation}
\label{eq5b}
\langle (\beta - \langle \beta \rangle )^2 \rangle^{1/2} < 20^{\circ}
\ \
\mbox{for}
\ 
\beta = \tan^{-1} \left( \frac{v}{U+u} \right) .
\end{equation}
The minimum value of $\langle (\beta - \langle \beta \rangle )^2 \rangle^{1/2}$ reflects the strength of the turbulence. It is about $14^{\circ}$ in our data.

We also exclude cases that are not consistent with the law of $\langle w^2 \rangle / u_{\ast}^2 = b_w$ in Eq.~(\ref{eq2b}). The experimental value of $b_w$ lies between $1.3$ and $1.8$.\cite{ofsbta17,mmym17,zs07} Although the reason for such a discrepancy of $b_w$ is not yet known, it appears comparable to uncertainties of $b_w$ in observations.\cite{km06,hhs02} A criterion is thereby imposed as
\begin{equation}
\label{eq5c}
\frac{\langle w^2(z) \rangle}{u_{\ast}^2} < 1.8 .
\end{equation}
\end{subequations}
The excluded cases are likely where the wind velocity has varied too largely or the flow has been affected by some obstacle lying in the windward direction.

\begin{figure}[tbp]
\resizebox{8.0cm}{!}{\includegraphics*[3.1cm,8.6cm][17.7cm,27.2cm]{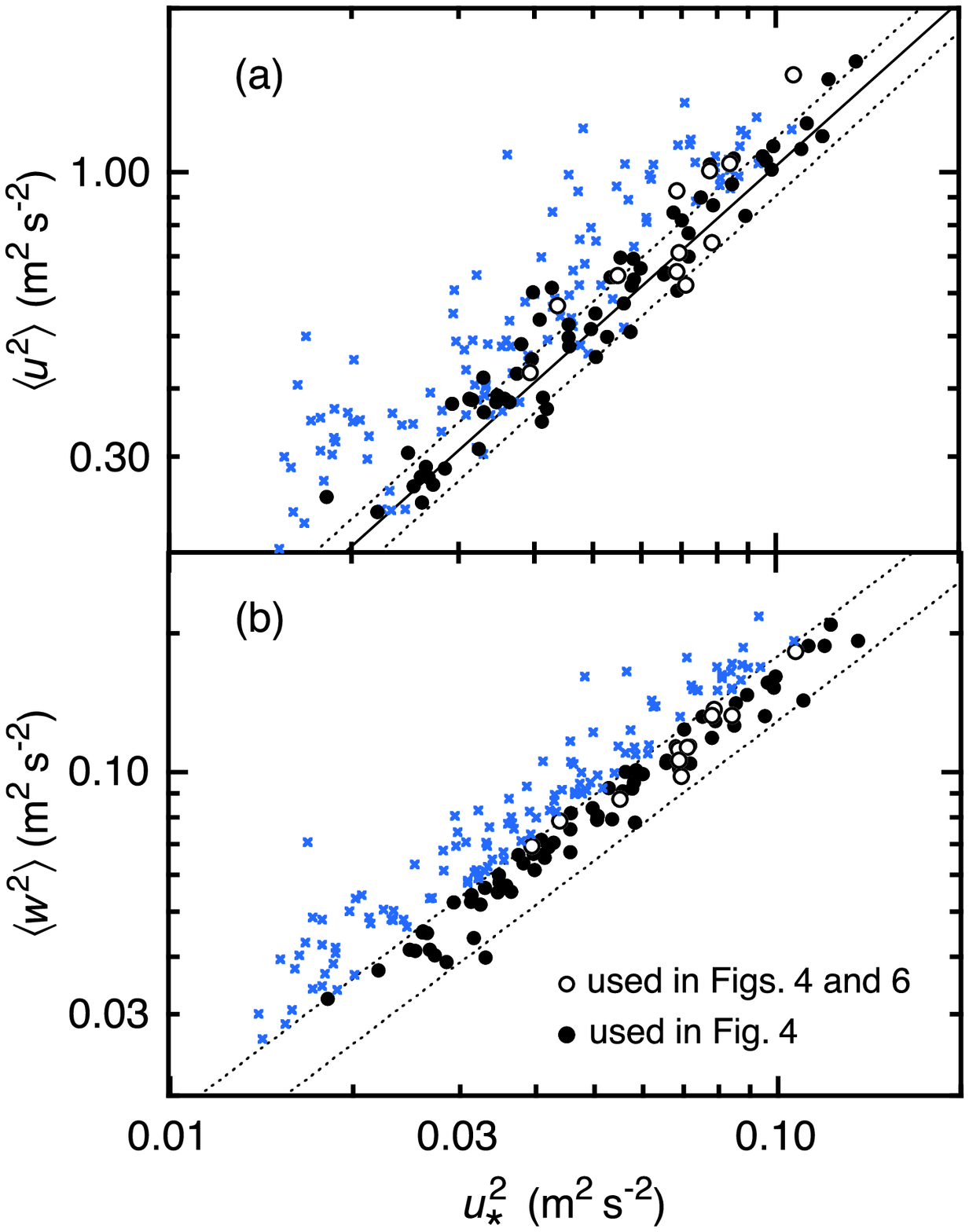}}
\caption{\label{f3} Comparison of $\langle u^2 \rangle$ and $\langle w^2 \rangle$ with $u_{\ast}^2$ for $11$ subsets of our data used in Figs.~\ref{f4} and \ref{f6} (open circles), $68$ subsets used in Fig.~\ref{f4} but not in Fig.~\ref{f6} (filled circles), and $156$ subsets not used in Figs.~\ref{f4} and \ref{f6} (crosses). The lines indicate $\langle u^2 \rangle /u_{\ast}^2 = 10.3 \pm 1.3$ (a) and $\langle w^2 \rangle /u_{\ast}^2 = 1.3$ and $1.8$ (b).}
\end{figure} 

Thus, $79$ subsets have been selected as a homogeneous sample of turbulence that is stationary, is near-neutral at least at the observing height of $z = 1.75$\,m, and does not suffer from precipitation or from any obstacle (see also Sec.~\ref{S6c}). While the viscosity $\nu$ is from $1.31 \times 10^{-5}$ to $1.44 \times 10^{-5}$\,m$^2$\,s$^{-1}$, the ratio $\langle u^2 \rangle /U^2$ is from $0.061$ to $0.118$.

We compare $\langle u^2 \rangle$ and $\langle w^2 \rangle$ with $u_{\ast}^2$ in Fig.~\ref{f3} (circles). The values of $\langle u^2 \rangle /u_{\ast}^2$ are consistent with $b_u + c_u \ln (\delta/z)$ in Eq.~(\ref{eq2a}) for $\delta = 1,000$\,m and $z = 1.75$\,m (solid line), if uncertainties of our data are comparable to $\pm 3\sigma$ errors (dotted lines) for $b_u = 2.30 \pm 0.18$ and $c_u = 1.26 \pm 0.06$ in a laboratory boundary layer.\cite{mmhs13} Our estimate $\delta = 1,000$\,m is a value typical of the total thickness of the near-neutral atmospheric boundary layer, $\delta \simeq 0.3 u_{\ast}/\vert f_{\rm C} \vert$ for the Coriolis parameter $f_{\rm C} = 1.46 \times 10^{-4} \sin \varphi$ in units of rad s$^{-1}$ at the latitude $\varphi$.\cite{g92,hhs02} This is analogous to $\delta_{99}$ in a laboratory boundary layer, i.e., a distance at which $U$ is $99$\% of its free-stream value (see Sec.~\ref{S6d}).

There are $156$ subsets that have been excluded with use of Eq.~(\ref{eq5c}). We also show them in Fig.~\ref{f3} (crosses). Since their values of $\langle u^2 \rangle /u_{\ast}^2$ and $\langle w^2 \rangle /u_{\ast}^2$ are simultaneously enhanced, it is confirmed that Eq.~(\ref{eq5c}) has served as a reliable criterion.

Finally, for the $79$ subsets selected here, the logarithmic law of $U(z) = (u_{\ast}/\kappa) \ln (z/z_0)$ is used to estimate the aerodynamic roughness length $z_0$ as $21 \pm 1$\,mm. Through a relation between the aerodynamic and the actual roughness lengths,\cite{g92} this estimate of $z_0$ is consistent with the grass height of our observing field, i.e., $\lesssim 50$\,mm (Sec.~\ref{S2}). Judging also from results in Fig.~\ref{f3}, the observing height of $z = 1.75$\,m is certainly within the constant-flux sublayer.

\section{Data Synthesis} \label{S4}

By using 79 subsets of our data selected in Sec.~\ref{S3}, we synthesize the energy spectra. Here and also in Sec.~\ref{S6b}, statistics for the whole data are distinguished from those for the individual subsets by numbering them as $m = 1$ to $M$ ($= 79$).

To $N = 2^{14}$ records of $X = u$ or $w$ in the $m$th subset, we apply cosine tapering of the edges of the $2 \times 2^4$ records. Any de-trending is not applied because it might modify the shape of the spectrum\cite{hhs02} and because non-stationary cases have been excluded in Sec.~\ref{S3}. From those tapered records, we obtain the Fourier transforms $\tilde{X}_{(m)}$ and the spectral densities ${\mit\Phi}_{X(m)} \varpropto \vert \tilde{X}_{(m)} \vert^2$.

We use Taylor's hypothesis to convert the frequencies into the wavenumbers, i.e., $k_{x(m,n)} = 2\pi f_{\rm s} n /NU_{(m)}$ at $n = 1$ to $N/2$. The mean velocity $U_{(m)}$ varies among the subsets $m$ in a range from $1.5$ to $4.5$\,m\,s$^{-1}$ (Sec.~\ref{S3}). Since the sampling frequency $f_s$ is $10$\,Hz and the distance $z$ is $1.75$\,m (Sec.~\ref{S2}), we have $k_{x(m,n)}z \simeq 4 \times 10^{-3}$ to $4 \times 10^1$ for $U_{(m)} = 1.5$\,m\,s$^{-1}$ and $k_{x(m,n)}z \simeq 1 \times 10^{-3}$ to $1 \times 10^1$ for $U_{(m)} = 4.5$\,m\,s$^{-1}$.

To synthesize the spectrum that would represent the whole data, $k_x {\mit\Phi}_{X(m)}/u_{\ast (m)}^2$ is averaged at each $k_x z$ over the subsets from $m=1$ to $M$. This is in accordance with Eqs.~(\ref{eq1a}) and (\ref{eq1b}), where $k_x {\mit\Phi}_X/u_{\ast}^2$ depends only on $k_x z$ at least at $k_x \ge a_1/\delta$. Although $a_1/\delta_{(m)}$ is not certain and is not constant among the subsets $m$, our approach is sufficient to examine the existence or nonexistence of the $k_x^{-1}$ law of Eq.~(\ref{eq1a}). If some other model were to be examined, we might require a different approach.

The actual synthesis is as follows. Since the distance $z$ has been fixed at $1.75$\,m, we consider the wavenumber $k_x$ alone. If $k_x$ is not too high and is not too low, most of the subsets have a pair of adjacent wavenumbers such that $k_{x(m,n)} \le k_x < k_{x(m,n+1)}$. Through a linear interpolation between $\ln [k_{x(m,n)}]$ and $\ln [k_{x(m,n+1)}]$, we calculate $k_x {\mit\Phi}_{X(m)}$ at that wavenumber $k_x$. Then, $k_x {\mit\Phi}_{X(m)}/u_{\ast (m)}^2$ is averaged over those subsets to estimate $k_x {\mit\Phi}_X/u_{\ast}^2$. Its uncertainty is estimated statistically from the variance of $k_x {\mit\Phi}_{X(m)}/u_{\ast (m)}^2$ in a standard manner.

\begin{figure}[bp]
\resizebox{8.0cm}{!}{\includegraphics*[3.1cm,8.6cm][17.7cm,27.2cm]{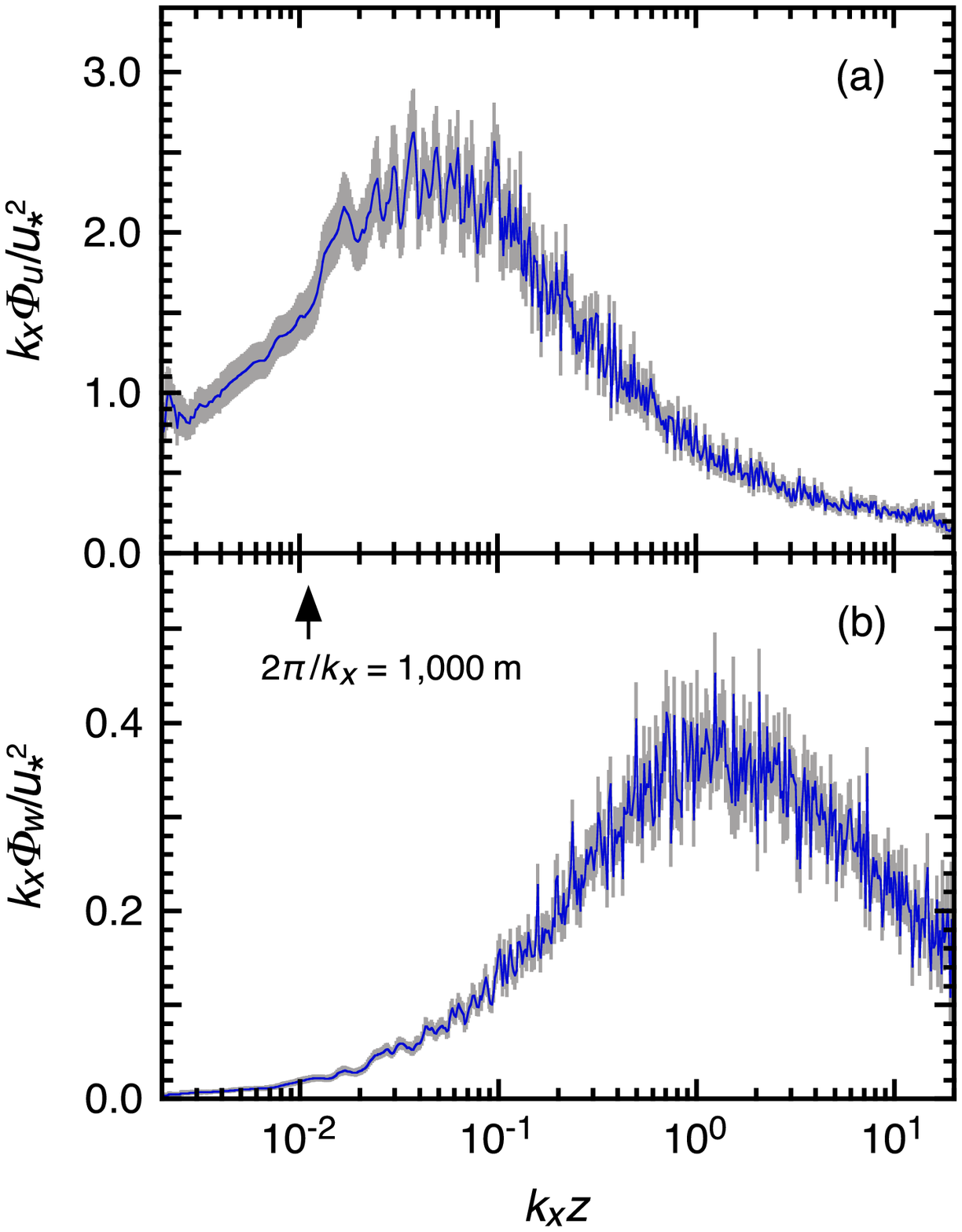}}
\caption{\label{f4} Premultiplied spectra $k_x {\mit\Phi}_u/u_{\ast}^2$ (a) and $k_x {\mit\Phi}_w/u_{\ast}^2$ (b) as a function of $k_x z$ for $79$ subsets of our data. The error bars of $\pm 1\sigma$ are also given (gray lines). An arrow indicates a wavenumber for $2\pi /k_x = 1,000$\,m at $z = 1.75$\,m.}
\end{figure} 

\section{Results} \label{S5}

The normalized densities of the premultiplied spectra $k_x {\mit\Phi}_u/u_{\ast}^2$ and $k_x {\mit\Phi}_w/u_{\ast}^2$ are shown as a function of $k_x z$ in Fig.~\ref{f4}. They represent wall turbulence at a Reynolds number $\delta u_{\ast}/\nu$ of $\mathcal{O}(10^7)$ and with the ratio $\delta /z$ of $\mathcal{O}(10^3)$ if we take values from Sec.~\ref{S3}. Compared with those in the laboratory experiments,\cite{zs07,rhvbs13,vgs15} $\delta u_{\ast}/\nu$ and $\delta/z$ are both enhanced by a factor of $\gtrsim 10$.

Being in contrast to most observations of the atmospheric surface layer,\cite{ff99,hhs02,km06,chosp13,wz16,kc98} we have not smoothed the spectra. They are noisy (see also Sec.~\ref{S6b}), but the overall shapes are evident.

First, we examine $k_x {\mit\Phi}_w$ in Fig.~\ref{f4}(b). As has been assumed to derive Eq.~(\ref{eq1b}), $k_x {\mit\Phi}_w$ is significant at $k_x z \gtrsim 10^0$ and is not at $k_x z \lesssim 10^{-1}$. These shape and magnitude of $k_x {\mit\Phi}_w$ are close to those in the experiments\cite{zs07} and in the previous observations.\cite{ff99,dckbfr04,hhs02,km06} We are hence able to rely on our data.

Then, we examine $k_x {\mit\Phi}_u$ in Fig.~\ref{f4}(a). Its magnitude is close to that of $k_x {\mit\Phi}_w$ at $k_x z \gtrsim 10^0$. With a decrease in $k_x z$, while $k_x {\mit\Phi}_w$ decreases, $k_x {\mit\Phi}_u$ increases still more. It exhibits a broad peak at $k_x z \simeq$ several times $10^{-2}$. This wavenumber is yet higher than $k_x = 2 \pi /\delta$ (arrow), i.e., $k_x$ for the total thickness of the near-neutral atmospheric boundary layer $\delta = 1,000$\,m (Sec.~\ref{S3}). Between $k_x \simeq \delta^{-1}$ and $z^{-1}$, there is not found the $k_x^{-1}$ law or the plateau of $k_x {\mit\Phi}_u / u_{\ast}^2 = c_u$. The peak value of $k_x {\mit\Phi}_u/u_{\ast}^2$ is larger than $c_u \simeq 1.3$, an estimate through Eq.~(\ref{eq2a}) from the variance $\langle u^2 \rangle$.\cite{mmhs13,hvbs12,hvbs13,mm13,vhs15,mmym17,ofsbta17}

Since Taylor's hypothesis has been used in Sec.~\ref{S4}, its effect is discussed here. Near the wall, low-wavenumber modes are advected faster than the mean velocity $U$ at the observing distance $z$ from the wall.\cite{dj09} These are due to largest eddies. With wall-normal sizes comparable to the turbulence thickness $\delta$, their advection is determined by some average of $U$ over distances from $z$ to $\delta$.\cite{t76,phc86} It is possible that Taylor's hypothesis has redistributed such low-wavenumber modes into the higher wavenumbers and has modified the shape of the peak of $k_x {\mit\Phi}_u$ in Fig.~\ref{f4}(a). Nevertheless, the peak itself is certain to exist because it is significant enough.\cite{rhvbs13} The observed ratios of $\langle u^2 \rangle /U^2$ are also not too large, i.e., $\lesssim 0.1$ (Sec.~\ref{S3}), as a condition for a use of Taylor's hypothesis.

The existence of such a peak of $k_x {\mit\Phi}_u$ is consistent with the well known existence of a minimum of $k_x {\mit\Phi}_u$ between meteorological variations at lower wavenumbers and the turbulence of the atmospheric boundary layer.\cite{my75,llp16} Usually, this has been studied on the basis of the frequencies. In a near-neutral and near-surface case,\cite{hhs02} $k_x {\mit\Phi}_u$ is minimal at several times $10^{-4}$\,Hz. It corresponds to $k_x z$ in-between $10^{-3}$ and $10^{-2}$ if the mean velocity $U$ is about $3$\,m\,s$^{-1}$ at our observing height of $z = 1.75$\,m.

Thus, our long observation does not support the $k_x^{-1}$ law found in some short observations.\cite{ff99,hhs02,chosp13,dckbfr04} The shape of $k_x {\mit\Phi}_u$ in Fig.~\ref{f4}(a) is rather close to those obtained from experiments of boundary layers and pipe flows.\cite{zs07,rhvbs13,vgs15} In each case, $k_x {\mit\Phi}_u$ exhibits a peak at $k_x z \simeq$ several times $10^{-2}$. Our peak is the highest. This result is explainable by Eq.~(\ref{eq3}), a prediction of the attached-eddy hypothesis. Its functional form is in accordance with the manner of our data synthesis (see Sec.~\ref{S4}), through which its factor $\ln (\delta /z)$ could lead to a large value representative of the atmospheric surface layer. The wavenumber of that peak is reproduced if the function $g_{u:2}(k_x z)$ is maximal there (see Fig.~\ref{f1}). We also consider that the other function $g_{u:1}(k_x z)$ is maximal at $k_x z \simeq$ several times $10^{-1}$, where those experiments have captured a shoulder of $k_x {\mit\Phi}_u$ if the ratio $\delta/z$ is not very large.\cite{zs07,rhvbs13,vgs15}

\section{Discussion} \label{S6}

\subsection{Reason for nonexistence of the $k_x^{-1}$ law} \label{S6a} 

To justify our result for nonexistence of the $k_x^{-1}$ law, the Fourier transform $\tilde{u}(k_x,z)$ of the velocity field of a single eddy $u(x,z)$ is studied at some distance $z$ from the wall. Since the total thickness of the wall turbulence $\delta$ is finite, we assume that the streamwise size of this eddy is finite and is up to $\mathcal{O}(\delta)$ as in the case of the attached-eddy hypothesis.\cite{t76}

From a mathematical theorem for compact supports in Sec.~\ref{S1},\cite{s66,b85} it follows that $\tilde{u}(k_x,z)$ could be nonzero at any of $k_x$. To explain this, the eddy is assumed to have one edge at $x = 0$ and the other at $x > 0$. At around $x = 0$, the velocity field is described with use of an integer $j \ge 1$ and of constants $c_0 \ne 0$, $c_1$, $c_2$, ... as
\begin{subequations}
\label{eq6}
\begin{equation}
\label{eq6a}
{\everymath{\displaystyle}
u(x,z) = \left\{ \begin{array}{ll}
                                                     0                                                 &\ \ \mbox{at} \ x <   0 ,  \\
                                  \rule{0ex}{3.5ex}  x^j \left( c_0 + c_1 x + c_2 x^2 +... \right)     &\ \ \mbox{at} \ x \ge 0 .
                 \end{array}
         \right.
}
\end{equation}
The other edge has a similar velocity field. By repeating a partial integration $j$ times\cite{s66} and by noting that $u(x,z)$ and its derivatives are equal to $0$ outside of that eddy,
\begin{equation}
\label{eq6b}
{\everymath{\displaystyle}
\begin{array}{ll}
\tilde{u}(k_x,z)                    & = \frac{1}{2 \pi}        \! \int_{-\infty}^{+\infty} \! \! \! u(x,z)       \exp(-ik_x x) dx \\
                 \rule{0ex}{4.0ex}  & = \frac{1}{2 \pi (ik)^j} \! \int_{-\infty}^{+\infty} \! \! \! u^{(j)}(x,z) \exp(-ik_x x) dx .
\end{array}
}
\end{equation}
\end{subequations}
The $j$th derivative $u^{(j)}(x,z)$ has a discontinuity of $\varpropto c_0$ at $x = 0$. Because of the resultant Gibbs phenomenon, we have $\int_{-\infty}^{+\infty} \! u^{(j)}(x,z) \exp(-ik_x x) dx$ of $\mathcal{O}(k_x^{-1})$. Even if $k_x$ is very high, $\tilde{u}(k_x,z)$ persists as $\mathcal{O}(k_x^{-j-1})$.

There is no wavenumber $k_x$ that corresponds to the size of the above eddy. For such a finite-size eddy, we might prefer a study in the real space, e.g., based on a two-point correlation rather than on the energy spectrum.

If wall turbulence is a random superposition of eddies, any effect of the total thickness $\delta$ is through the largest eddies of a streamwise size of $\varpropto \delta$. According to Eq.~(\ref{eq6b}), such an effect is persistent at any high wavenumber. This is in contrast to a condition for the $k_x^{-1}$ law of Eq.~(\ref{eq1a}), which requires that $\delta$ affects low-wavenumber modes but not the high-wavenumber modes. Actually, the $k_x^{-1}$ law is inconsistent with Eq.~(\ref{eq3}) and hence with the attached-eddy hypothesis.\cite{m17} This is also true for eddies not attached to the wall so far as the streamwise size of any of them is finite.

The same discussion holds for Fourier modes of finite-size eddies in the two- or three-dimensional wavenumber space. We do not expect any law that is to correspond to the one-dimensional $k_x^{-1}$ law.

\subsection{Fluctuations of the spectral density} \label{S6b}

The distribution of an instantaneous spectral density ${\mit\Phi}_{X(m)} = {\mit\Phi}_{u(m)}$ or ${\mit\Phi}_{w(m)}$ is studied among independent subsets of data from $m=1$ to $M$. They are assumed to have been obtained from the same turbulence. We define ${\mit\Phi}_{X}$ as $ \sum_{m=1}^M {\mit\Phi}_{X(m)}/M$ in the limit $M \rightarrow +\infty$.

Within the constant-flux sublayer of wall turbulence, fluctuations of $u$ and $w$ are closely Gaussian.\cite{ff96,mthk03,mm13,vhs15} We thereby model the turbulence as a homogeneous Gaussian random field,\cite{my71,my75} where the fluctuations of $X = u$ or $w$ are exactly Gaussian. The Fourier transforms $\tilde{X}$ are also Gaussian and are independent of one another. Their statistics are determined by ${\mit\Phi}_X$ alone.

The random field of this type is from the central limit theorem.\cite{my71} For example, in the attached-eddy hypothesis, the eddies are randomly superposed on one another in space.\cite{t76} With an increase in their number density, those fluctuations are increasingly Gaussian.\cite{m17,mm13}

At any wavenumber $k_x$ of the $m$th subset of such data, the real and the imaginary parts of the Fourier transform $\tilde{X}_{(m)}$ fluctuate independently in a common zero-mean Gaussian distribution.\cite{my75,mthk02} In turn, ${\mit\Phi}_{X(m)} \varpropto \mbox{Re}[\tilde{X}_{(m)}]^2 + \mbox{Im}[\tilde{X}_{(m)}]^2$ fluctuates in a $\chi^2$ distribution with $2$ degrees of freedom, i.e., in an exponential distribution,
\begin{subequations}
\label{eq7}
\begin{equation}
\label{eq7a}
P_1 \left( \phi \right) = \exp \left( - \phi \right)
\ \
\mbox{for}
\ 
\phi = \frac{{\mit\Phi}_{X(m)}}{{\mit\Phi}_X} \ge 0 .
\end{equation}
The probability density $P_1$ is maximal not at $\phi = \langle \phi \rangle = 1$ but at $\phi = 0$. It leads to $\langle (\phi - \langle \phi \rangle)^2 \rangle^{1/2} = 1$. Thus, ${\mit\Phi}_{X(m)}$ could differ significantly from ${\mit\Phi}_X$. Even if smoothed over nearby wavenumbers, the difference remains over those where the Fourier transforms are sparse. As pointed out in Sec.~\ref{S1}, discrepancies are actually found among spectra obtained from short observations.\cite{ff99,hhs02,dckbfr04,km06,chosp13,wz16} Those mimicking the $k_x^{-1}$ law would be an extreme example.

By averaging ${\mit\Phi}_{X(m)}$ over the subsets from $m = 1$ to $M$, we obtain an estimate of ${\mit\Phi}_X$. Its fluctuation is described by a $\chi^2$ distribution with $2M$ degrees of freedom,
\begin{equation}
\label{eq7b}
P_M \left( \phi \right) = \frac{\phi^{M-1} e^{- M\phi}}{(M-1)!/M^M}
\ \
\mbox{for}
\ 
\phi = \frac{\sum_{m =1}^M \! {\mit\Phi}_{X(m)}/M}{{\mit\Phi}_X} \ge 0.
\end{equation}
\end{subequations}
While $P_M$ is maximal at $\phi = 1-M^{-1}$ that tends to $\phi = \langle \phi \rangle = 1$ as $M \rightarrow +\infty$, we have $\langle (\phi - \langle \phi \rangle)^2 \rangle^{1/2} = M^{-1/2}$. For the present case of $M = 79$, the result of $M^{-1/2} = 11$\,\% is comparable to magnitudes of noisy fluctuations of the spectral densities in Fig.~\ref{f4}. They are dominated certainly by the turbulence itself.

\subsection{Data selection thresholds and spectrum} \label{S6c}

We have selected our data with use of thresholds in Sec. \ref{S3}. To confirm that their values are sufficient for the final result of $k_x {\mit\Phi}_u/u_{\ast}^2$ in Fig.~\ref{f4}(a), they are independently strengthened as $-25^{\circ}$ instead of $-40^{\circ}$ and $-10^{\circ}$ instead of $+5^{\circ}$ for the mean wind direction $\alpha$ in Eq.~(\ref{eq4a}), $0.02$ instead of $0.1$ for the significance of the buoyancy $\vert z/L_{\ast} \vert$ in Eq.~(\ref{eq5a}), $17^{\circ}$ instead of $20^{\circ}$ for the standard deviation of the wind direction $\langle (\beta - \langle \beta \rangle )^2 \rangle^{1/2}$ in Eq.~(\ref{eq5b}), or $1.6$ instead of $1.8$ for the normalized variance of the vertical wind velocity $\langle w^2 \rangle /u_{\ast}^2$ in Eq.~(\ref{eq5c}). Here, our data used in Fig.~\ref{f4}(a) are at $\langle (\beta - \langle \beta \rangle )^2 \rangle^{1/2} \ge 14^{\circ}$ and at $\langle w^2 \rangle /u_{\ast}^2 \ge 1.2$. The total number of the subsets $M$ is $22$, $34$, $27$ or $30$, respectively, instead of $79$.

\begin{figure}[tbp]
\rotatebox{90}{\resizebox{5.7cm}{!}{\includegraphics*[2.0cm,4.8cm][18.5cm,28.0cm]{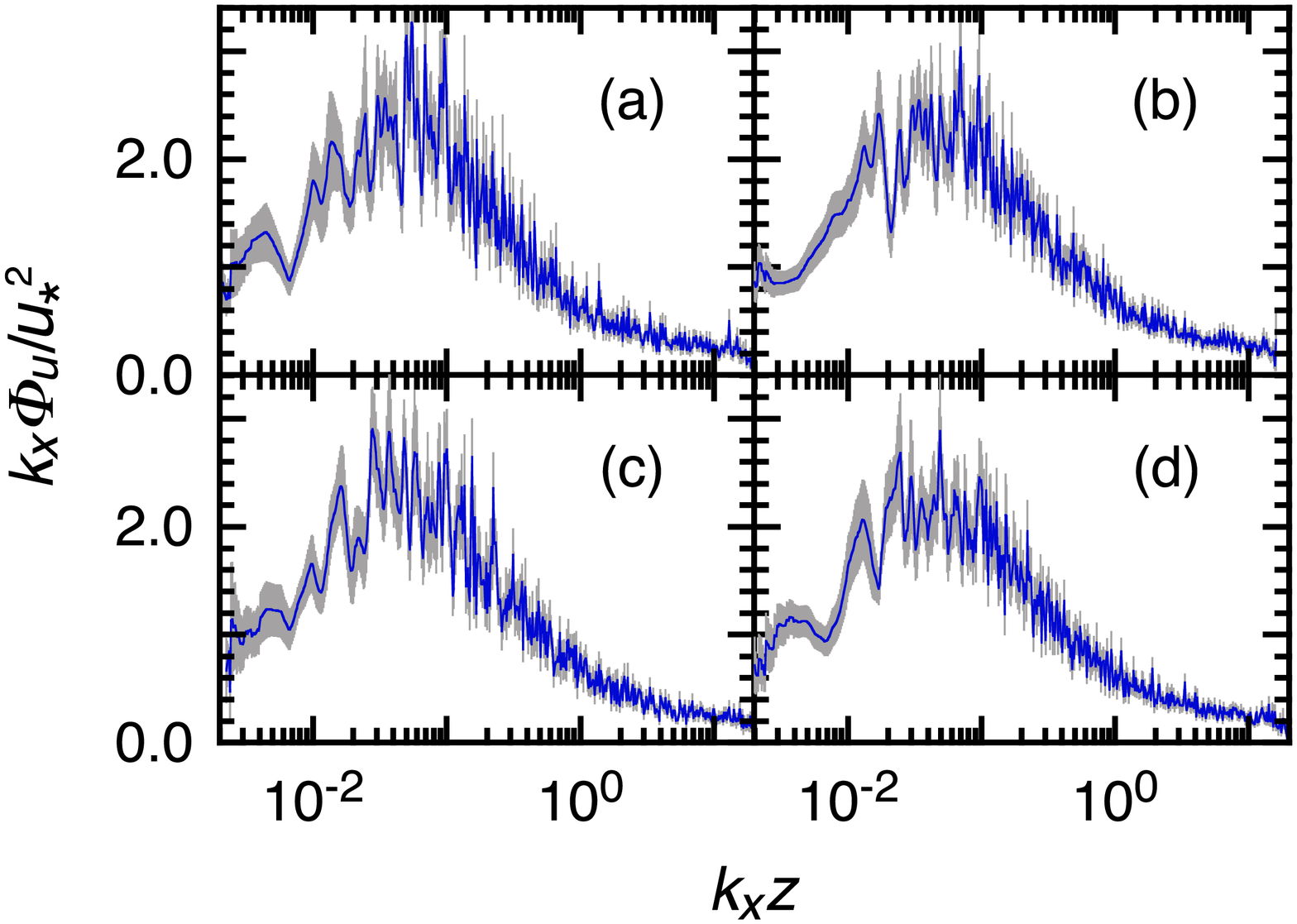}}}
\caption{\label{f5} Premultiplied spectrum $k_x {\mit\Phi}_u/u_{\ast}^2$ as a function of $k_x z$ for (a) $22$ subsets of our data at $-25^{\circ} \le \alpha < -10^{\circ}$, (b) $34$ subsets at $\vert z/L_{\ast} \vert < 0.02$, (c) $27$ subsets at $\langle (\beta - \langle \beta \rangle )^2 \rangle^{1/2} < 17^{\circ}$, and (d) $30$ subsets at $\langle w^2 \rangle /u_{\ast}^2 < 1.6$ out of those used in Fig.~\ref{f4}. The error bars of $\pm 1\sigma$ are also given (gray lines).}
\end{figure}

The resultant spectra are shown in Fig.~\ref{f5}. Although noisy fluctuations are large because the subset number $M$ is small (see Sec.~\ref{S6b}), they are not inconsistent with that in Fig.~\ref{f4}(a).

\subsection{Thickness of the atmospheric boundary layer} \label{S6d}

The turbulence thickness $\delta$ is equated to $\delta_{99}$ in the case of a laboratory boundary layer. As for $\delta$ in the atmospheric boundary layer, there is a controversy. While the thickness of the surface layer of $\mathcal{O}(100\,\mbox{m})$ has been used in some studies,\cite{wz16,km06} the total thickness of $\mathcal{O}(1,000\,\mbox{m})$ has been used here and in other studies.\cite{hhs02,dckbfr04,kc98} We justify our use of $\mathcal{O}(1,000\,\mbox{m})$.

The nearby Aerological Observatory (Sec.~\ref{S3}) is a site for routine launches of radio sondes (Meisei, RS-11G or iMS-100). With a height resolution of $5$ to $10$\,m, a sonde measures the velocity $v_{\rm sonde}$ and the direction $\alpha_{\rm sonde}$ of the instantaneous horizontal wind. The pressure and the temperature are also measured so as to obtain the potential temperature $\theta_{\rm sonde}$.\cite{g92} Among the $79$ subsets of our data used in Fig.~\ref{f4}, we find $11$ subsets within $\pm 30$\,min of the center time of the sonde observation, $0000$ or $1200$ UT. These observations are examined here.

The individual profiles of $v_{\rm sonde}$, $\alpha_{\rm sonde}$, and $\theta_{\rm sonde}$ are shown in Fig.~\ref{f6}. We also indicate the heights corresponding to our estimates of the total thickness of the boundary layer, i.e., $0.3 u_{\ast}/\vert f_{\rm C} \vert$ in Sec.~\ref{S3} (circles). Up to each of these heights of $z \simeq 1,000$\,m, while $v_{\rm sonde}$ tends to increase, $\alpha_{\rm sonde}$ and $\theta_{\rm sonde}$ tend to remain constant. Such features are attributable to fully developed turbulence of a near-neutral boundary layer. Even above $z \simeq 1,000$\,m, $v_{\rm sonde}$ and $\alpha_{\rm sonde}$ are very variable, since two-dimensional and long-wavelength motions exist there. At the largest heights, a west wind is prevailing as is usual in the middle latitudes.

Thus, between $\delta_{99}$ in a laboratory boundary layer and the total thickness of $\mathcal{O}(1,000\,\mbox{m})$ in the near-neutral atmospheric boundary layer, there is an analogy. Although $v_{\rm sonde}$ tends to exhibit a local maximum at $z < 1,000$\,m in Fig.~\ref{f6}(a), this is because $v_{\rm sonde}$ reflects both the mean and the fluctuating velocities.

As for the surface layer with thickness of $\mathcal{O}(100\,\mbox{m})$, we have related it to the constant-flux sublayer (Sec.~\ref{S1}). This is in accordance with the meteorology literature.\cite{g92,hhs02,dckbfr04,kc98} Actually in data obtained from a tall tower that had been located at $200$\,m from our observing field until several years ago, $\langle -uw \rangle$ is almost constant and $U$ is almost logarithmic up to a height of $z \simeq 150$\,m.\cite{hhao12}

Finally, we note that our estimates $\delta = 0.3 u_{\ast}/\vert f_{\rm C} \vert$ are based on our data of $u_{\ast} = \langle -uw \rangle^{1/2}$. Since they reproduce satisfactorily the observations of the total thickness of the boundary layer in Fig.~\ref{f6}, our data are reliable despite having been obtained under non-optimal conditions at a single height of $z = 1.75$\,m (Sec.~\ref{S2}).

\section{Concluding Remarks} \label{S7}

For the constant-flux sublayer of wall turbulence, the existence or nonexistence of the $k_x^{-1}$ law of Eq.~(\ref{eq1a}) has been examined observationally. We had continued a field observation of the atmospheric surface layer over several months, have selected data with Eq.~(\ref{eq5}), and have used them to synthesize the spectral density ${\mit\Phi}_u$. The result of $k_x {\mit\Phi}_u$ in Fig.~\ref{f4}(a) is representative of wall turbulence at a high Reynolds number of $\delta u_{\ast}/\nu \simeq 10^7$ and with a large ratio of $\delta /z \simeq 10^3$. It is not consistent with the $k_x^{-1}$ law of Eq.~(\ref{eq1a}) and is rather consistent with Eq.~(\ref{eq3}), a prediction of the attached-eddy hypothesis.

The underlying assumption for the $k_x^{-1}$ law is that low-wavenumber modes are independent of the distance from the wall $z$ whereas high-wavenumber modes are independent of the total thickness of the turbulence $\delta$.\cite{pa77} Over the overlapping wavenumbers, there is the $k_x^{-1}$ law. This does not hold if the wall turbulence is a superposition of finite-size eddies as has been assumed in the attached-eddy hypothesis.\cite{t76} For such a case, Eq.~(\ref{eq6b}) implies that those high-wavenumber modes are also dependent on the largest eddies and on their size of $\varpropto \delta$.

\begin{figure}[tbp]
\rotatebox{90}{\resizebox{6.1cm}{!}{\includegraphics*[4.7cm,6.0cm][19.7cm,27.1cm]{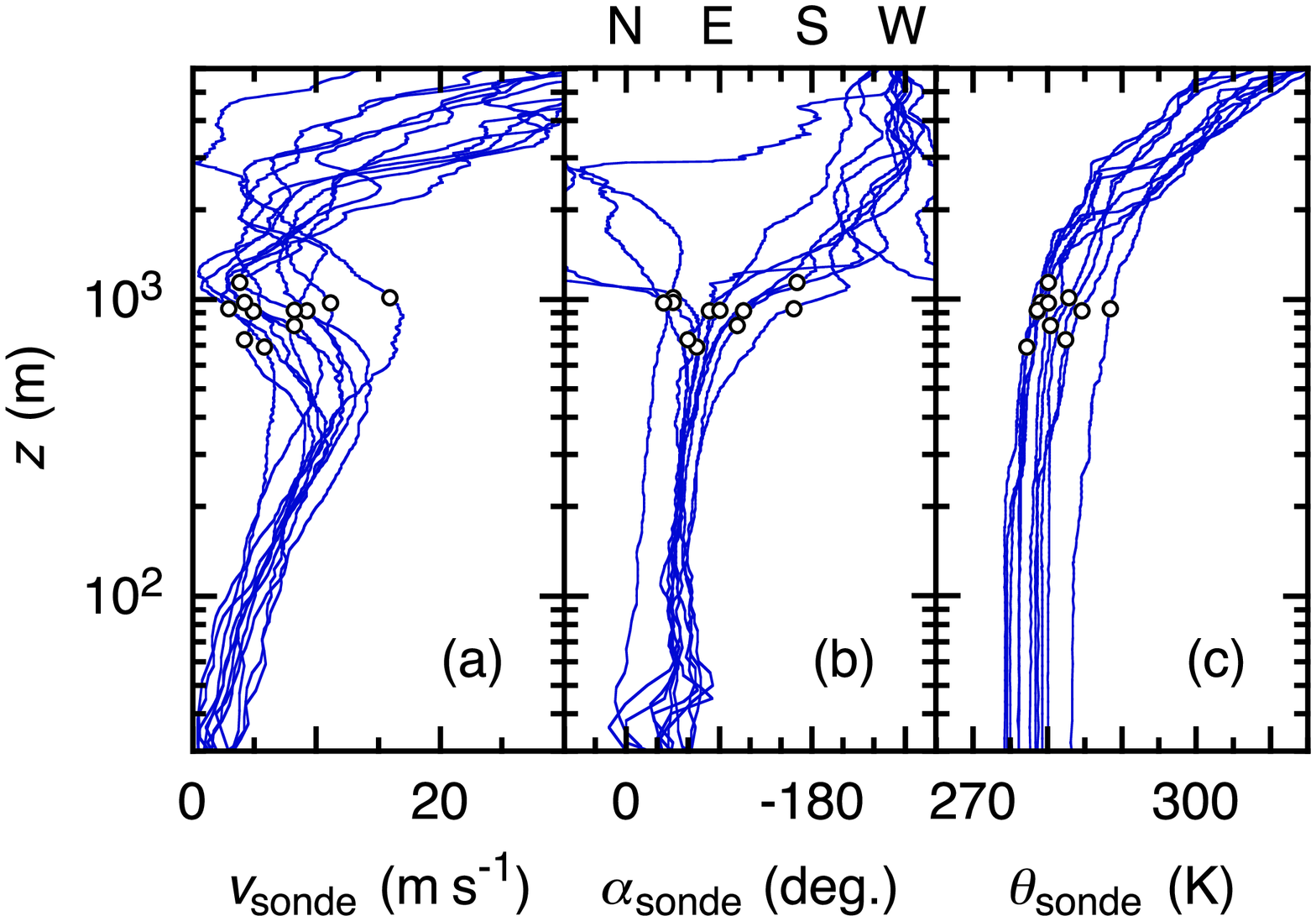}}}
\caption{\label{f6} Profiles of (a) $v_{\rm sonde}$, (b) $\alpha_{\rm sonde}$, and (c) $\theta_{\rm sonde}$ as a function of $z$ for $11$ subsets of our data out of those used in Fig. \ref{f4}. The circles indicate the heights corresponding to estimates of the total thickness of the boundary layer, $0.3 u_{\ast}/\vert f_{\rm C} \vert$.}
\end{figure}

Since obstacles lie around our observing field (Fig.~\ref{f2}), we have excluded any data that appear affected by them. The remaining data are explainable consistently as those of a boundary layer with total thickness of $\delta \simeq 1,000$\,m (Figs.~\ref{f3} and \ref{f6}), i.e., a typical case in the near-neutral atmosphere. It is still crucial to confirm our result with observational data obtained under optimal conditions like those at dry lake beds.\cite{km06,wz16,kc98} Subsets of the data have to be as many as ours, according to Eq.~(\ref{eq7b}) that describes statistical convergence of the spectral density.

The logarithmic law of a variance appears to exist also for fluctuations of the pressure\cite{jh08} and of the concentration of a passive scalar.\cite{mmym17} To study their energy spectra at a large ratio of $\delta/z$, a field observation would be useful as in our study of $k_x {\mit\Phi}_u$.

The $k_x^{-1}$ law for Eq.~(\ref{eq1a}) as well as the attached eddy for Eq.~(\ref{eq3}) are no more than approximate models and are not to capture each feature of the turbulence. In particular, wall turbulence is known to contain very large structures with streamwise lengths that are at least a few times of its total thickness $\delta$.\cite{a07} They are made of substructures that appear attached to the wall. We might have to consider, e.g., alignment of the attached eddies up to a length of $\varpropto \delta$.\cite{a07} Nevertheless, those structures are meandering. The one-dimensional Fourier modes are likely to be dominated not by the structures themselves but by their substructures. Actually in Fig.~\ref{f4}(a), the peak of $k_x {\mit\Phi}_u$ lies at $k_x > 2 \pi / \delta$ so far as $\delta$ is $\mathcal{O}(1,000\,\mbox{m})$. The wavenumber of $2 \pi / \delta$ is rather close to that for the known minimum of $k_x {\mit\Phi}_u$.\cite{hhs02} Thus, before considering the details, it is important to establish which of the $k_x^{-1}$ law,\cite{pa77,phc86,t53} the attached eddy,\cite{t76,m17} or some other model is most suited to approximating the constant-flux sublayer of wall turbulence at high Reynolds numbers.

\begin{acknowledgments}

This work was supported in part by KAKENHI Grant No.~17K00526. We are grateful to the Aerological Observatory of the Japan Meteorological Agency for the sonde data and so on.

\end{acknowledgments}


\begin{thebibliography}{999}

\bibitem{my71} A.\,S. Monin and A.\,M. Yaglom, {\it Statistical Fluid Mechanics} (MIT Press, Cambridge, 1971), Vol. 1.

\bibitem{g92} J.\,R. Garratt, {\it The Atmospheric Boundary Layer} (Cambridge University Press, Cambridge, U.K., 1992).

\bibitem{pa77} A.\,E. Perry and C.\,J. Abell, ``Asymptotic similarity of turbulence structures in smooth- and rough-walled pipes,'' J. Fluid Mech. {\bf 79}, 785--799 (1977).

\bibitem{phc86} A.\,E. Perry, S. Henbest, and M.\,S. Chong, ``A theoretical and experimental study of wall turbulence,'' J. Fluid Mech. {\bf 165}, 163--199 (1986).

\bibitem{t53} C.\,M. Tchen, ``On the spectrum of energy in turbulent shear flow,'' J. Res. Natl. Bur. Stand. {\bf 50}, 51--62 (1953). 

\bibitem{hvbs12} M. Hultmark, M. Vallikivi, S.\,C.\,C. Bailey, and A.\,J. Smits, ``Turbulent pipe flow at extreme Reynolds numbers,'' Phys. Rev. Lett. {\bf 108}, 094501 (2012).

\bibitem{mmhs13} I. Marusic, J.\,P. Monty, M. Hultmark, and A.\,J. Smits, ``On the logarithmic region in wall turbulence,'' J. Fluid Mech. {\bf 716}, R3 (2013).

\bibitem{hvbs13} M. Hultmark, M. Vallikivi, S.\,C.\,C. Bailey, and A.\,J. Smits, ``Logarithmic scaling of turbulence in smooth- and rough-wall pipe flow,'' J. Fluid Mech. {\bf 728}, 376--395 (2013).

\bibitem{mm13} C. Meneveau and I. Marusic, ``Generalized logarithmic law for high-order moments in turbulent boundary layers,'' J. Fluid Mech. {\bf 719}, R1 (2013).

\bibitem{vhs15} M. Vallikivi, M. Hultmark, and A.\,J. Smits, ``Turbulent boundary layer statistics at very high Reynolds number,'' J. Fluid Mech. {\bf 779}, 371--389 (2015).

\bibitem{ofsbta17} R. \"Orl\"u, T. Fiorini, A. Segalini, G. Bellani, A. Talamelli, and P.\,H. Alfredsson, ``Reynolds stress scaling in pipe flow turbulence\,---\,first results from CICLoPE,'' Phil. Trans. R. Soc. A {\bf 375}, 20160187 (2017). 

\bibitem{mmym17} H. Mouri, T. Morinaga, T. Yagi, and K. Mori, ``Logarithmic scaling for fluctuations of a scalar concentration in wall turbulence,'' Phys. Rev. E {\bf 96}, 063101 (2017).

\bibitem{t76} A.\,A. Townsend, {\it The Structure of Turbulent Shear Flow}, 2nd ed. (Cambridge University Press, Cambridge, U.K., 1976).

\bibitem{m17} H. Mouri, ``Two-point correlation in wall turbulence according to the attached-eddy hypothesis,'' J. Fluid Mech. {\bf 821}, 343--357 (2017).

\bibitem{s66} L. Schwartz, {\it Mathematics for the Physical Sciences} (Hermann, Paris, France, 1966).

\bibitem{b85} M. Benedicks, ``On Fourier transforms of functions supported on sets of finite Lebesgue measure,'' J. Math. Anal. Appl. {\bf 106}, 180--183 (1985). 

\bibitem{lm15} M. Lee and R.\,D. Moser, ``Direct numerical simulation of turbulent channel flow up to $Re_{\tau} \approx 5200$,'' J. Fluid Mech. {\bf 774}, 395--415 (2015).

\bibitem{zs07} R. Zhao and A.\,J. Smits, ``Scaling of the wall-normal turbulence component in high-Reynolds-number pipe flow,'' J. Fluid Mech. {\bf 576}, 457--473 (2007).

\bibitem{rhvbs13} B.\,J. Rosenberg, M. Hultmark, M. Vallikivi, S.\,C.\,C. Bailey, and A.\,J. Smits, ``Turbulence spectra in smooth- and rough-wall pipe flow at extreme Reynolds numbers,'' J. Fluid Mech. {\bf 731}, 46--63 (2013).

\bibitem{vgs15} M. Vallikivi, B. Ganapathisubramani, and A.\,J. Smits, ``Spectral scaling in boundary layers and pipes at very high Reynolds numbers,'' J. Fluid Mech. {\bf 771}, 303--326 (2015).

\bibitem{wz16} G. Wang and X. Zheng, ``Very large scale motions in the atmospheric surface layer: a field investigation,'' J. Fluid Mech. {\bf 802}, 464--489 (2016).

\bibitem{km06} G.\,J. Kunkel and I. Marusic, ``Study of the near-wall-turbulent region of the high-Reynolds-number boundary layer using an atmospheric flow,'' J. Fluid Mech. {\bf 548}, 375--402 (2006).

\bibitem{hhs02} U. H\"ogstr\"om, J.\,C.\,R. Hunt, and A.-S. Smedman, ``Theory and measurements for turbulence spectra and variances in the atmospheric neutral surface layer,'' Boundary-Layer Meteorol. {\bf 103}, 101--124 (2002). 

\bibitem{dckbfr04} P. Drobinski, P. Carlotti, R.\,K. Newsom, R.\,M. Banta, R.\,C. Foster, and J.-L. Redelsperger, ``The structure of the near-neutral atmospheric surface layer,'' J. Atmos. Sci. {\bf 61}, 699--714 (2004). 

\bibitem{ff99} P.\,L. Fuehrer and C.\,A. Friehe, ``A physically-based turbulent velocity time series decomposition,'' Boundary-Layer Meteorol. {\bf 90}, 241--295 (1999). 

\bibitem{chosp13} M. Calaf, M. Hultmark, H.\,J. Oldroyd, V. Simeonov, and M.\,B. Parlange, ``Coherent structures and the $k^{-1}$ spectral behaviour,'' Phys. Fluids {\bf 25}, 125107 (2013).

\bibitem{kc98} G. Katul and C.-R. Chu, ``A theoretical and experimental investigation of energy-containing scales in the dynamic sublayer of boundary-layer flows,'' Boundary-Layer Meteorol. {\bf 86}, 279--312 (1998). 

\bibitem{dj09} J.\,C. del \'Alamo and J. Jim\'enez, ``Estimation of turbulent convection velocities and corrections to Taylor's approximation,'' J. Fluid Mech. {\bf 640}, 5--26 (2009).

\bibitem{my75} A.\,S. Monin and A.\,M. Yaglom, {\it Statistical Fluid Mechanics} (MIT Press, Cambridge, 1975), Vol. 2.

\bibitem{llp16} X.\,G. Lars\'en, S.\,E. Larsen, and E.\,L. Petersen, ``Full-scale spectrum of boundary-layer winds,'' Boundary-Layer Meteorol. {\bf 159}, 349--371 (2016). 

\bibitem{ff96} H.\,H. Fernholz and P.\,J. Finley, ``The incompressible zero-pressure-gradient turbulent boundary layer: an assessment of the data,'' Prog. Aerosp. Sci. {\bf 32}, 245--311 (1996). 

\bibitem{mthk03} H. Mouri, M. Takaoka, A. Hori, and Y. Kawashima, ``Probability density function of turbulent velocity fluctuations in a rough-wall boundary layer,'' Phys. Rev. E {\bf 68}, 036311 (2003).

\bibitem{mthk02} H. Mouri, M. Takaoka, A. Hori, and Y. Kawashima, ``Probability density function of turbulent velocity fluctuations,'' Phys. Rev. E {\bf 65}, 056304 (2002).

\bibitem{hhao12} M. Horiguchi, T. Hayashi, A. Adachi, and S. Onogi, ``Large-scale turbulence structures and their contributions to the momentum flux and turbulence in the near-neutral atmospheric boundary layer observed from a $213$\,m tall meteorological tower,'' Boundary-Layer Meteorol. {\bf 144}, 179--198 (2012). 

\bibitem{jh08} J. Jim\'enez and S. Hoyas, ``Turbulent fluctuations above the buffer layer of wall-bounded flows,'' J. Fluid Mech. {\bf 611}, 215--236 (2008).

\bibitem{a07} R.\,J. Adrian, ``Hairpin vortex organization in wall turbulence,'' Phys. Fluids {\bf 19}, 041301 (2007).

\end{thebibliography}
\end{document}